# On optical rotation and selective transmission in ambichiral sculptured thin films

Ferydon Babaei

Department of Physics, University of Qom, Qom, Iran
Email: fbabaei@qom.ac.ir

**Abstract**

The optical rotation and selective transmission spectrum of ambichiral sculptured thin films using transfer matrix method have been calculated. The results of optical modeling showed that optical characteristics ambichiral sculptured thin films are the same as chiral sculptured thin films in lower angular rotations. In higher angular rotations appear two circular Bragg regimes. The primary at shorter wavelengths reflects LCP light and the secondary at longer wavelengths reflects RCP light. The optical properties of ambichiral sculptured thin films with twist and layer defects as spectral holes in circular Bragg regimes have been reported.

Keywords: Sculptured thin films; Optical rotation

## I. Introduction

Chiral sculptured thin films (CSTFs) are three dimensional structures that may be produced by using oblique angle deposition combined with the rotation of substrate about its normal to surface axis [1]. The intriguing feature in CSTFs is due to appearance of circular Bragg regime (or circular Bragg phenomenon) in these structures [2]. The utilizing of the circular Bragg regime, CSTFs as circular polarization filters have been theoretically examined and then experimentally realized [2, 3]. By introducing either a layer defect or a twist defect in the middle of a CSTF, narrowband spectral–hole filters have been designed and implemented [4-6].

The ambichiral sculptured thin films (ACSTFs) are a stack of biaxial plates that can be produced by combination of oblique angle deposition and abruptly changes of substrate rotation about its surface normal [7].The discrete angular rotations create two circular Bragg regimes in these films. Then, in fabrication of CSTFs, substrate rotation as continuously or discretely will affect the circular Bragg regimes and also adjusting defects in order to excitation spectral holes in Bragg regimes is significant.

In this work the optical rotation and selective transmission of an ACSTF have been calculated and compared with the results that obtained of a CSTF as a reference structure. Theory in brief outlined in section II. Numerical results are presented and discussed in Section III.

## II. Theory in brief

Consider a region $0 \leq z \leq d$ be occupied by a CSTF or ACSTF (Fig.1). While the regions $z \leq 0$ and $z \geq d$ are vacuous. Let the structure be excited by a plane wave propagating at an angle $\theta_{inc}$ to the z- axis and at an angle $\psi_{inc}$ to the x- axis in the xy-plane. The phasors of incident, reflected and transmitted electric fields are given as [8]:

$$\begin{cases} \underline{E}_{inc}(\underline{r}) = [(\frac{i\underline{S}-\underline{P}_+}{\sqrt{2}})a_L - (\frac{i\underline{S}+\underline{P}_+}{\sqrt{2}})a_R]e^{i\underline{k}_0 \cdot \underline{r}} , & z \leq 0 \\ \underline{E}_{ref}(\underline{r}) = [-(\frac{i\underline{S}-\underline{P}_-}{\sqrt{2}})r_L + (\frac{i\underline{S}+\underline{P}_-}{\sqrt{2}})r_R]e^{-i\underline{k}_0 \cdot \underline{r}} , & z \leq 0 \\ \underline{E}_{tr}(\underline{r}) = [(\frac{i\underline{S}-\underline{P}_+}{\sqrt{2}})t_L - (\frac{i\underline{S}+\underline{P}_+}{\sqrt{2}})t_R]e^{i\underline{k}_0 \cdot (\underline{r}-d\underline{u}_z)} , & z \geq d \end{cases} \quad (1)$$

The magnetic field's phasor in any region is given as $\underline{H}(\underline{r}) = (i\omega\mu_0)^{-1}\underline{\nabla} \times \underline{E}(\underline{r})$ where $(a_L, a_R), (r_L, r_R)$ and $(t_L, t_R)$ are the amplitudes of incident plane wave, and reflected and transmitted waves with left- or right-handed polarizations. We also have:

$$\begin{cases} \underline{r} = x\underline{u}_x + y\underline{u}_y + z\underline{u}_z \\ \underline{k}_0 = k_0(\sin\theta_{inc}\cos\psi_{inc}\underline{u}_x + \sin\theta_{inc}\sin\psi_{inc}\underline{u}_y + \cos\theta_{inc}\underline{u}_z) \end{cases} \quad (2)$$

The unit vectors for linear polarization normal and parallel to the incident plane, $\underline{S}$ and $\underline{P}$, respectively are defined as:

$$\begin{cases} \underline{S} = -\sin\psi_{inc}\underline{u}_x + \cos\psi_{inc}\underline{u}_y \\ \underline{P}_\pm = \mp(\cos\theta_{inc}\cos\psi_{inc}\underline{u}_x + \cos\theta_{inc}\sin\psi_{inc}\underline{u}_y) + \sin\theta_{inc}\underline{u}_z \end{cases} \quad (3)$$

The reflectance and transmittance amplitudes can be obtained, using the continuity of the tangential components of electrical and magnetic fields at two interfaces $z = 0$, $d$ and solving the algebraic matrix equation [8]:

$$\begin{bmatrix} i(t_L - t_R) \\ -(t_L + t_R) \\ 0 \\ 0 \end{bmatrix} = \left[\underline{\underline{K}}(\theta_{inc},\psi_{inc})\right]^{-1} \cdot \left[\underline{\underline{B}}(d,\Omega)\right] \cdot \left[\underline{\underline{M}}'(d,\Omega,\kappa,\psi_{inc})\right] \cdot \left[\underline{\underline{K}}(\theta_{inc},\psi_{inc})\right] \cdot \begin{bmatrix} i(a_L - a_R) \\ -(a_L + a_R) \\ -i(r_L - r_R) \\ (r_L + r_R) \end{bmatrix} \quad (4)$$

The different terms and parameters of this equation are given in detail (see equations (2-25), (2-26) in reference [8]. Selective transmission of the structure (CSTF or ACSTF) is $T_{LL} - T_{RR}$ and the reflection and transmission can be calculated as

$$R_{i,j} = \left|\frac{r_i}{a_j}\right|, \quad T_{i,j} = \left|\frac{t_i}{a_j}\right|^2 \; ; \; i,j = L, R.$$

The optical rotation is the angle through which the major axis of the transmission vibration ellipse has rotated with respect to the major axis of the incidence vibration ellipse [9]. The tilt angles between the unit vector $\underline{S}$ and the major axes of the vibration ellipse of the incident and transmitted electric field vectors are determined using a procedure detailed by Chen [10]:

$$\begin{aligned} \tau_{inc} &= \tan^{-1}\left(\frac{|a_R|\operatorname{Re}[a_L] + |a_L|\operatorname{Re}[a_R]}{|a_R|\operatorname{Im}[a_L] - |a_L|\operatorname{Im}[a_R]}\right) \\ \tau_{tr} &= \tan^{-1}\left(\frac{|t_R|\operatorname{Re}[t_L] + |t_L|\operatorname{Re}[t_R]}{|t_R|\operatorname{Im}[t_L] - |t_L|\operatorname{Im}[t_R]}\right) \end{aligned} \quad (5)$$

The optical rotation is then defined as [10, 11]:

$$\phi_{tr} = \begin{cases} \tau_{tr} - \tau_{inc} + \pi, & \text{if} \quad -\pi \leq \tau_{tr} - \tau_{inc} \leq -\pi/2 \\ \tau_{tr} - \tau_{inc}, & \text{if} \quad |\tau_{tr} - \tau_{inc}| \leq \pi/2 \\ \tau_{tr} - \tau_{inc} - \pi, & \text{if} \quad \pi/2 \leq \tau_{tr} - \tau_{inc} \leq \pi \end{cases} \quad (6)$$

It is possible that the value of calculated $\phi_{tr}$ becomes $|\phi_{tr}| > \pi/2$. In this case, if $\phi_{tr} > \pi/2$ or $\phi_{tr} < -\pi/2$ then by subtracting or adding a value of $\pi$, respectively, the $\phi_{tr}$ value will be confined in the range $[-\pi/2, \pi/2]$.

### III. Numerical results and discussion

Consider a right-handed sculptured thin film (CSTF or ACSTF) in its bulk state has occupied the free space. The relative permittivity scalars $\varepsilon_{a,b,c}$ in this sculptured thin film were obtained using the Bruggeman homogenization formalism [12]. In this formalism, the film is considered as a two component composite (vacuum and material deposition). These quantities are dependent on different parameters, namely, columnar form factor, fraction of vacuum (void fraction ($f_v$)), the wavelength of free space and the refractive index $n(\lambda_0) + ik(\lambda_0)$ of the film's material (inclusion). In addition, each column in the STF structure is considered as a string of identical long ellipsoids [13]. In all our calculations, columnar form factors $\gamma_\tau^s = \gamma_\tau^v = 20$, $\gamma_b^s = \gamma_b^v = 1.11$ (s and v, respectively indicate to inclusion and vacuum phase) and structural parameters of CSTF $\chi = 42°$, $2\Omega = 325nm$, $f_v = 0.421$ were fixed [14]. In order to obtain the optical rotation and selective transmission spectrum in the axial excitation ($\theta_{inc} = \psi_{inc} = 0^0$), we have used the experimental data of the dielectric refractive index bulk titania (TiO$_2$) [15]. Also, we have included the dispersion and dissipation of dielectric function.

The optical rotation(for P-polarization) and selective transmission for an ACSTF have depicted in Fig.2 at different abrupt angular rotations .We have increased angular rotation from 5° to 120° with a step 5° and optical spectra few of them are plotted in Fig.2. Until $\varphi = 70^0$, optical spectra of ACSTF similar results of those of CSTF as a reference structure are obtained. A CSTF contains a single circular Bragg band centered at the Bragg wavelength $\lambda^{Br} \approx 2 n_{avg} \Omega$, where $n_{avg}$ is the average refractive index. However, a ACSTF contains two circular Bragg bands , a primary circular Bragg band at $\lambda^{Br}$ and an inverted circular Bragg band centered at $\lambda^{inv} \approx \frac{\varphi}{180^0 - \varphi} \lambda^{Br}$, where $\varphi$ is the angular rotation of the substrate between each arm of the ACSTF (in degrees)[7].

Our calculations showed that the intensity of secondary circular Bragg band is negligible until $\varphi = 70^0$ and occurs at shorter wavelengths below 300nm. At $\varphi = 70^0$ inverted circular Bragg band appears about 440nm with 54% maximum selective transmittance and with increasing angular rotation shifts to longer wavelengths. For a tetragonal CSTF two circular Bragg bands occur at $\lambda^{inv} \approx \lambda^{Br}$ and obtained a residual selective transmission near to zero.

In order to understand the effect of angular rotation, a ACSTF can be described as a stack of biaxial plates that rotate by $\varphi°$ increments in the clockwise direction as seen by RCP light and by (180- $\varphi$ )° increments in the counterclockwise direction as seen by LCP light [7]. The different rotation rates are responsible for the formation of two circular Bragg bands. We can divide each revolution to $n = \frac{360}{\varphi}$ sectors. If the number of sectors to be high, the optical properties of ACSTF is similarly to a CSTF, otherwise, two circular Bragg bands are observed.

The optical properties of trigonal CSTF have been compared with a reference structure in the presence a central 90°- twist and spacing layer defect in Fig.3. By inserting a twist defect in the half thickness of the structure creates a 180° phase shift, placing the spectral hole at the center of circular reflection bands [13]. The occurrence of spectral holes for trigonal CSTF in our work is at wavelengths 618nm, 1195nm. Then a phase discontinuity in the structure of ACSTFs will produce spectral holes in the two circular reflection bands. Also, the optical rotation (for P-polarization) and selective transmission by adding a spacing layer defect $d_{spa} = 150\,nm$ in half thickness of trigonal CSTF in compare with a CSTF have been depicted in Fig.3. We considered the relative permittivity scalars $\varepsilon_{a,b,c}$ of layer defect as $\varepsilon_a^{spa} = \varepsilon_b^{spa} = \varepsilon_c^{spa} = \dfrac{\varepsilon_c + \varepsilon_d}{2}$ where

$$\varepsilon_d = \dfrac{\varepsilon_a \varepsilon_b}{\varepsilon_a \cos^2 \chi + \varepsilon_b \sin^2 \chi}$$ [13]. Then, by adding layer defect to structure, placing a spectral hole roughly in the middle of the Bragg regime. The layer defect affects primary circular reflection band in trigonal CSTF with splitting them as spectral hole. The occurrence of spectral hole is located at 606nm. The optical thickness would give a $2\pi$ phase shift at the wavelength of the primary reflection band, and thus not create a spectral hole in inverted reflection band. The results achieved in this work are consistent with the experimental data (Popta *et al.* (2005)).The calculations for optical rotation were repeated for a S-polarization plane wave and the same results obtained.

## IV. Conclusions

In this work, we theoretically analyzed the optical rotation and selective transmission of ACSTFs using transfer matrix method. In compare to CSTFs, the results showed that in the low angular rotations ($\varphi < 70^0$) do not exist difference between optical properties of ACSTFs and CSTFs. However, the difference appears at $\varphi = 70^0$ as two circular reflection bands. Because in an ACSTF (as stack of biaxial plates) a number of biaxial plates oriented in the clockwise direction ($\varphi°$ rotation rate) and the rest in the counterclockwise direction (($180-\varphi$)° rotation rate). The different rotation rates can to be a reason for the creation of two circular Bragg bands. Spectral holes, by adding twist and layer defects to structure of ACSTFs, appear in two circular reflection bands. The results of this work may be applied to the optical characterization of chiral sculptured thin films.

**Acknowledgements**

We wish to acknowledge support from the University of Qom.

**Figure captions**

Fig.1. Schematic of a single column of (a) a CSTF, (b) a trigonal CSTF, (c) the top-view of (a, b), (d) a trigonal CSTF with a central twist defect , (e) a trigonal CSTF with a spacing layer defect. The pitch of structure is $2\Omega$ and $\chi$ is the angle of rise.

Fig.2. (a) Optical rotation of linearly polarized light (P-polarization), (b) selective transmittance of circularly polarized light in right-handed CSTF and ACSTF. The structure is described by the following parameters $\chi = 42°, 2\Omega = 325\,nm$ and $d = 30\Omega$ .

Fig.3. (a) Optical rotation of linearly polarized light (P-polarization), (b) selective transmittance of circularly polarized light in right-handed CSTF and trigonal CSTF with a central $90°$ - twist and a spacing layer defect. The half- thicknesses of CSTF and trigonal CSTF with a defect are respectively considered as $d = 15\Omega$ and $d = 30\Omega$ .The spacing layer is described by the parameters $d_{spa} = 150\,nm$ and $\varepsilon_a^{spa} = \varepsilon_b^{spa} = \varepsilon_c^{spa} = \dfrac{\varepsilon_c + \varepsilon_d}{2}$ .

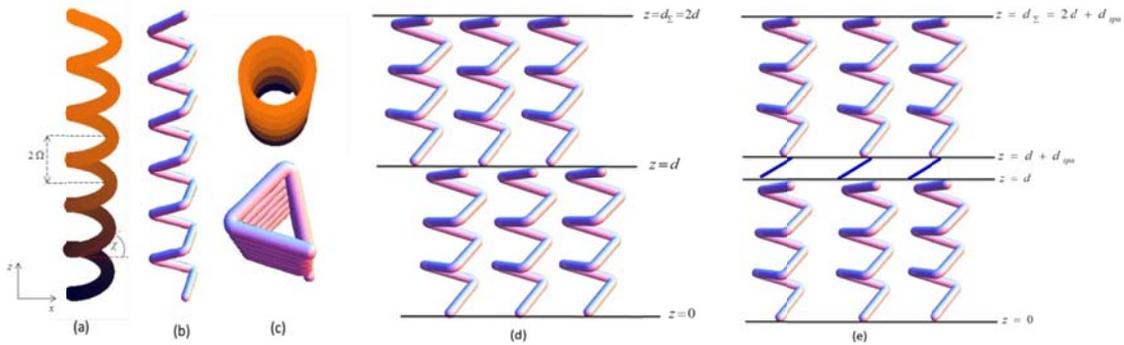

**Fig.1; F. Babaei**

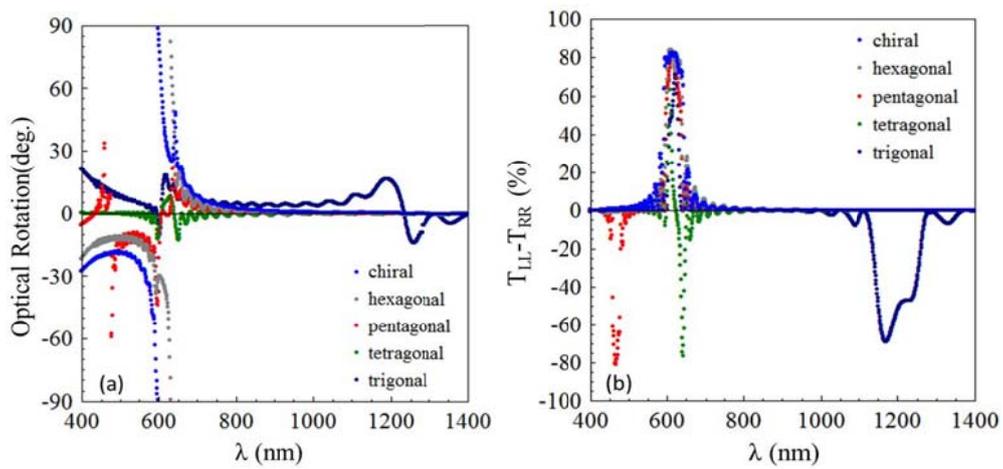

**Fig. 2; F. Babaei**

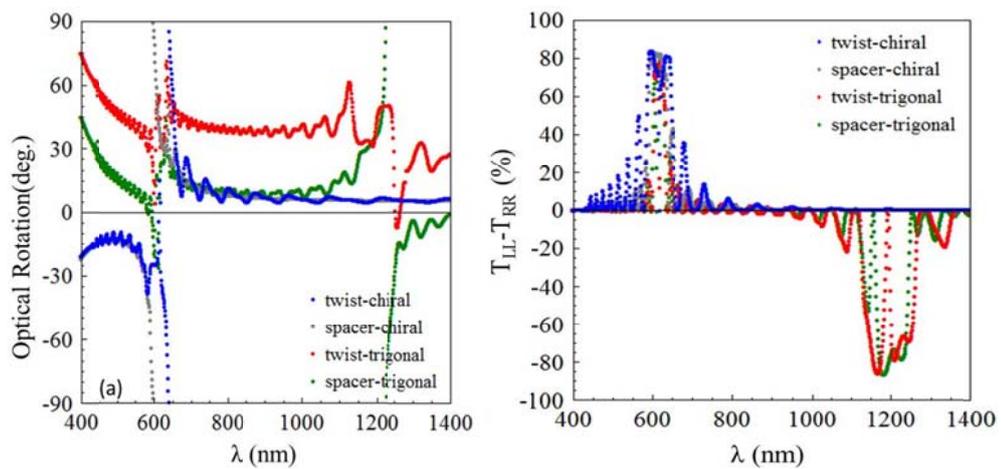

**Fig.3; F. Babaei**